\documentclass[review]{elsarticle}
\usepackage{geometry}
\usepackage{amsmath}
\usepackage{amsfonts}
\usepackage{amssymb}
\usepackage{graphicx}
\usepackage{dcolumn}
\usepackage{bm}
\usepackage[colorlinks = true, allcolors = blue]{hyperref}
\usepackage{float}
\usepackage{subcaption}

\makeatletter
\def\ps@pprintTitle{%
	\let\@oddhead\@empty
	\let\@evenhead\@empty
	\def\@oddfoot{}%
	\let\@evenfoot\@oddfoot}
\makeatother
\begin{document}
\definecolor{background-color}{gray}{0.98}

	\newcommand{\bea}{\begin{eqnarray}}
	\newcommand{\eea}{\end{eqnarray}}
	\newcommand{\bes}{\begin{subequations}}
	\newcommand{\ees}{\end{subequations}}
\newtheorem{dfn}{Definition}[section]
\newtheorem{ex}{Example}[section]
\newtheorem{subex}{Example}[subsection]
\newtheorem{cl}{Corrolary}[section]
\newtheorem{propo}{Proposition}[section]
\newtheorem{theorem}{Theorem}[section]

\newcommand{\bd}{\begin{document}}
\newcommand{\ed}{\end{document}}
\newcommand{\bc}{\begin{center}}
\newcommand{\ec}{\end{center}}
\newcommand{\bfr}{\begin{flushright}}
\newcommand{\efr}{\end{flushright}}
\newcommand{\lt}{\left}
\newcommand{\rt}{\right}
\newcommand{\vs}{\vspace}
\newcommand{\hs}{\hspace}
\newcommand{\beq}{\begin{equation}}
\newcommand{\eeq}{\end{equation}}
\newcommand{\lb}{\linebreak}
\newcommand{\pb}{\pagebreak}
\newcommand{\mb}{\makebox}
\newcommand{\fb}{\framebox}
\newcommand{\mc}{\multicolumn}
\newcommand{\ben}{\begin{enumerate}}
\newcommand{\een}{\end{enumerate}}
\newcommand{\bit}{\begin{itemize}}
\newcommand{\eit}{\end{itemize}}
\newcommand{\oln}{\overline}
\newcommand{\un}{\underline}
\newcommand{\lefq}{\lefteqn}
\newcommand{\ba}{\begin{array}}
\newcommand{\ea}{\end{array}}
\newcommand{\beqa}{\begin{eqnarray}}
\newcommand{\eeqa}{\end{eqnarray}}
\newcommand{\beqas}{\begin{eqnarray*}}
\newcommand{\eeqas}{\end{eqnarray*}}
\newcommand{\bfg}{\begin{figure}}
\newcommand{\efg}{\end{figure}}
\newcommand{\bds}{\begin{displaymath}}
\newcommand{\eds}{\end{displaymath}}
\newcommand{\btb}{\begin{tabbing}}
\newcommand{\etb}{\end{tabbing}}
\newcommand{\para}{\parallel}
\newcommand{\pad}{\partial}
\newcommand{\nn}{\nonumber}
\newcommand{\la}{\leftarrow}
\newcommand{\ra}{\rightarrow}
\newcommand{\lgla}{\longleftarrow}
\newcommand{\lgra}{\longrightarrow}
\newcommand{\La}{\Leftarrow}\newcommand{\Ra}{\Rightarrow}
\newcommand{\Lra}{\Leftrightarrow}
\newcommand{\Lgla}{\Longleftarrow}
\newcommand{\Lgra}{\Longrightarrow}
\newcommand{\lan}{\langle}
\newcommand{\ran}{\rangle}
\renewcommand{\a}{\alpha}
\renewcommand{\b}{\beta}
\newcommand{\g}{\gamma}
\newcommand{\G}{\Gamma}
\renewcommand{\d}{\delta}
\newcommand{\eps}{\epsilon}
\newcommand{\Th}{\Theta}
\newcommand{\s}{\sigma}
\newcommand{\lam}{\lambda}
\newcommand{\D}{\Delta}
\newcommand{\ds}{\displaystyle}
\newcommand{\vare}{E}
\newcommand{\pr}{\prime}
\newcommand{\ro}{\rho}
\newcommand{\nab}{\nabla}
\newcommand{\m}{\mu}
\newcommand{\n}{\nu}
\newcommand{\Sg}{\Sigma}
\newcommand{\p}{\pi}
\newcommand{\R}{I\!\!R}
\newcommand{\om}{\omega}
\newcommand{\Om}{\Omega}
\newcommand{\ovra}{\overrightarrow}
\newcommand{\ze}{\zeta}
\newcommand{\vart}{\vartheta}
\newcommand{\tri}{\triangle}
\newcommand{\f}{\frac}
\newcommand{\iny}{\infty}
\newcommand{\pro}{\propto}
\renewcommand{\arraystretch}{1.25}

\begin{frontmatter}

\title{Ro-vibrational energy analysis of Manning-Rosen and P\"oschl-Teller potentials with a new improved approximation in the 
centrifugal term}

\author[address1]{Debraj Nath}
\ead{debrajn@gmail.com}

\author[address2]{Amlan K.~Roy}
\ead{akroy6k@gmail.com}

\address[address1]{Department of Mathematics, Vivekananda College, Kolkata-700063, WB, India.}
\address[address2]{Department of Chemical Sciences, Indian Institute of Science Education and Research (IISER) Kolkata, Mohanpur-741246, 
Nadia, WB, India.}

\begin{abstract}
Two physically important potentials (Manning-Rosen and P\"oschl-Teller) are considered for the ro-vibrational energy in diatomic molecules. 
An improved new approximation is invoked for the centrifugal term, which is then used for their solution within the Nikiforov-Uvarov framework.  
This employs a recently proposed scheme, which combines the two widely used Greene-Aldrich and Pekeris-type approximations. Thus, approximate 
analytical expressions are derived for eigenvalues and eigenfunctions. The energies are examined with respect to two approximation parameters, 
$\lam$ and $\nu$. The original approximations are recovered for certain specials values of these two parameters. This offers a simple effective 
scheme for these and other relevant potentials in quantum mechanics. 
 \\ \\
\end{abstract} 

\end{frontmatter}

\section{Introduction}
The construction of a general, universal potential energy function for molecules has been at the forefront of research 
activity in chemical, molecular and solid-state physics, as it carries the relevant information for a given molecule. 
They are needed as input in various areas including spectroscopy, molecule-molecule collision, molecular simulation, 
dynamical situation, thermodynamic properties etc. 
There are non-trivial challenges faced towards the development of energy-distance relationship, due to which, its general 
form still remains elusive. The publication of famous exponential Morse potential about 90 years ago, has inspired a broad
range of works along this direction having varying degrees of sophistication, accuracy and flexibility. A vast number of  
such functions have been generated over the years by a number of researchers. Generally speaking, as the number of parameters
in the analytic potential function increases, the greater it fits with the experimental data. 

The exponential Manning-Rosen (MR) potential \cite{Manning.Rosen},
\beq\label{pot.MR}
V=V_{MR}(r)=\f{\hbar^2}{2\mu b^2} \left[\f{\a(\a-1)e^{-2r/b}}{(1-e^{-r/b})^2}-\f{Ae^{-r/b}}{1-e^{-r/b}}\right],
\eeq
has played the role of an important mathematical model in the context of molecular vibration and rotation. It has relevance 
to a multitude of bound and resonance-state problems in physics. This expression contains two dimensionless parameters $A$ 
(signifying the strength) and $\alpha$ (constant), whereas the screening parameter $b$ (having a dimension of length) is 
connected to the range of potential. Some interesting properties include (i) it is invariant under transformation $\alpha 
\leftrightarrow \alpha -1$ (ii) this reduces to the celebrated short-range Hulth\'en potential for $\alpha=0$ or 1 (iii) a 
relative minimum occurs for $r=b \ln \left[ 1+ \frac{2 \alpha (\alpha -1)}{A} \right]$, when $\alpha > 1$ and $A > 0$. 

The Schr\"odinger equation for the so-called $s$ (non-zero $\ell$) states can be obtained \emph{exactly}, by a number of
elegant and attractive formalisms, such as direct factorization method, Feynman path-integral formalism, standard 
function analysis leading to wave functions expressed in terms of hyper-geometric functions, a tridiagonal matrix 
representation \cite{diaf2005,MR, chen2007}, etc. However, the same for $\ell \neq 0$ states are yet to be found in closed
analytic form, giving rise to a number of notable approximate schemes. Thus, arbitrary $\ell$ states were calculated by 
means of an approximation of centrifugal term: $\frac{1}{r^2} \approx \frac{1}{b^2} \frac{e^{-r/b}}{(1-e^{-r/b})^2}$ in 
short range \cite{qiang2007}, similar along the lines of familiar Pekeris scheme. In the literature, other methods 
are also available, such as, a super-symmetric shape invariance formalism along with a function analysis \cite{chen2009}, 
$1/r^2$ approximated by a term having 3 adjustable parameters \cite{MR3}, Nikiforov-Uvarov (NU) method with 
an approximation of the $1/r^2$ term \cite{MR4.SMIkhdair}, path-integral formalism approached by a Duru-Kleinert method 
\cite{diaf2011}. In \cite{hady2011}, the author used Laguerre and oscillator bases for tridiagonalization of relevant
Hamiltonian and a Gauss quadrature to calculate potential matrix elements. Other significant works are: J-matrix method 
\cite{nasser2013} with a proper expansion for $1/r^2$, numerical integrating procedure in MATHEMATICA \cite{lucha1999}, 
generalized pseudospectral method \cite{MR9}, etc. While most works focused on bound states, scattering states 
\cite{chen2007,nasser2013} were also considered. Moreover, properties such as oscillator strengths, multipole moments, 
transition probabilities for certain states were reported in the literature \cite{nasser2013}. It has been probed in 
higher dimension \cite{gu2011}. 
 
Another important diatomic molecular potential is the P\"oschl-Teller (PT) \cite{Poschl.Teller} like model, given by, 
\beq\label{pot.PT}
V=V_{PT}(r)=\f{\xi_1-\xi_2\cosh(\a r)}{\sinh^2(\a r)},
\eeq
where $r$ denotes the internuclear distance, $\xi_1, \xi_2$ are two parameters controlling the potential well, and $\alpha$
governs the range of potential. Its bound states for arbitrary quantum numbers were expressed in terms of hyper-geometric 
functions $_2 F_1 (a,b,c; z)$, by employing an approximation to $1/r^2$ \cite{PT}. One of its variants, the so-called  
Scarf potential was analyzed by expanding the $1/r^2$ term around the minimum equilibrium point \cite{qiang2010}. 
A comparison of Pekeris- as well as Greene-Aldrich-like approximations for the centrifugal term, in the context of ro-vibrational states, 
have been presented lately \cite{PT2.epjp2020, horchani2020}. The scattering states have found discussion \cite{pekeris.ap2, you2013} as well.

Thus it appears that there is a preponderance of approaches which involve a well-designed approximation for the centrifugal term. Broadly 
speaking, two major routes have gained popularity over the years, namely the Pekeris- \cite{Pekeris,pekeris.badawi,pekeris.ap3,pekeris.ap4} 
and Greene-Aldrich-type 
approximations, along with their several variations \cite{greene}. The primary objective of this work is to 
present an extension/modification of such schemes for the two potentials mentioned in Eqs.~\ref{pot.MR} and \ref{pot.PT}.  
The new proposed scheme has been recently applied successfully for energy and thermodynamic analysis of Deng-Fan molecular potential \cite{nath2021}. 
Towards this end, we will express the centrifugal term as a linear combination of two approximations which turns out (from the future
discussion) to offer an accurate alternative approximation to the centrifugal term, near the origin ($r=0$) and any other point ($r=r_0$). 
This new intuitive approximation is succinctly discussed as functions of $r$, $r_0$ and two approximating parameters, namely, $\lam$, $\nu$. 
The corresponding energy spectra generated from our current approximation are then critically compared with the available established 
approximations for the centrifugal term, for some representative states. This is done for both MR and PT potentials. 
 
The article is organized as follows. Section \ref{Sec.Exac}, at first, provides the formal solutions of Schr\"odinger equation within 
the NU method. The eigenvalues and eigenfunctions for MR and PT within the various approximations of centrifugal term, are derived 
in Secs.~\ref{Sec.Examples1} and \ref{Sec.Examples2} respectively. The associated energy spectra are analyzed in Sec.~\ref{Sec.Results}. 
Moreover, the effect of the approximation parameters $\lam$ and $\nu$ on the energy spectrum of MR and PT potentials are shown in 
Finally a few comments are made in Sec.~\ref{Sec.Conclusion}.



\section{Exact solution of the Schr\"odinger equation by Nikiforov-Uvarov Method}\label{Sec.Exac}
Let us consider the Schr\"odinger equation of a diatomic molecule,  
\beq\label{Schro.Eq}
-\f{\hbar^2}{2\mu}\nabla^2\psi+V({\bf r})\psi=E\psi,
\eeq 
in presence of MR ($V({\bf r}) = V_{MR}(r))$ or PT ($V({\bf r}) = V_{PT}(r))$ potential of Eq.~(\ref{pot.MR}) and 
Eq.~(\ref{pot.PT}) respectively. Here,  
\beq
\nabla^2=\f{1}{r^2}\f{\partial}{\partial r}\left(r^2\f{\partial}{\partial r}\right)+\f{1}{r^2\sin\theta}\f{\partial}{\partial \theta}\left(\sin\theta\f{\partial}{\partial \theta}\right)+\f{1}{r^2\sin^2\theta}\f{\partial^2}{\partial \phi^2},
\eeq
and $\mu$ is the reduced mass of a diatomic molecule, $r$ is the internuclear distance. The MR potential, $V_{MR}$ has a minimum at 
$r=b\ln\left[1+\f{2\a(\a-1)}{A}\right]$, whereas the same occurs at $r=\f{1}{\a}\tanh^{-1}\left(\sqrt{\f{2a}{b}}\right)$ in case of
$V_{PT}$. Let
\beq
\psi({\bf r})=\f{R(r)}{r}Y_l^m(\theta)\,e^{im\phi},
\eeq 
be the solution of Eq.~(\ref{Schro.Eq}). Then one obtains, 
\beq\label{Eq.R}
\f{d^2R}{dr^2}+\left[\f{2\mu}{\hbar^2}E-\f{2\mu }{\hbar^2}V(r)-\f{l(l+1)}{r^2}\right]R=0,
\eeq 
and
\beq\label{Ylm}
Y_{l}^{m}=\left[\f{(2l+1)(l-|m|)!}{4\pi(l+|m|)!}\right]^{\f{1}{2}}\,P_{l}^{|m|}(\cos\theta)\,e^{im\phi},
\eeq 
where $l(l+1)$ is the separation constant. To find the general solution of resulting radial Schr\"odinger equation obtained in 
Eq.~(\ref{Eq.R}), we will use a transformation $s=e^{-r/b}$ for MR and $s=\tanh^2(\a r/2)$ for PT potentials respectively. Moreover, 
we have considered two set of bases, \emph{viz.},  
(i) $\left\{1,\f{e^{-r/b}}{1-e^{-r/b}},\f{e^{-2r/b}}{(1-e^{-r/b})^2}\right\}$, for MR and 
(ii) $\left\{1,\f{1}{\sinh^2\a r},\f{\cosh \a r}{\sinh^2\a r}\right\}$, for PT potential, for which the centrifugal term is approximated 
in different forms, such as Greene-Aldrich \cite{greene} and Pekeris-type \cite{Pekeris}. Then the Schr\"odinger equation becomes a second-order 
differential equation of the form given below, 
\beq\label{Eq.NU.gen}
\f{d^2R}{ds^2}+\f{\widetilde{\tau}(s)}{\sigma(s)}\f{dR}{ds}+\f{\widetilde{\sigma}(s)}{\sigma^2(s)}R=0,
\eeq 
where $\widetilde{\tau}(s)$, $\sigma(s)$, $\widetilde{\sigma}(s)$ are polynomials in $s$ of degree one, two and two respectively. 
If we let, 
\beq
R(s)=\phi_1(s)\phi_2(s),
\eeq
be the solution of Eq.~(\ref{Eq.NU.gen}), then we obtain \cite{NU}, 
\beq\label{Eq.y}
\sigma(s)\phi_2^{''}(s)+\tau(s)\phi'_2(s)+\nu\, \phi_2(s)=0,
\eeq
and
\beq
\ba{l}
\phi_1(s)=\ds e^{\int\f{\pi(s)}{\sigma(s)}ds},
\ea
\eeq 
where
\beq
\ba{ll}\label{pi.sigma}
\pi(s)&=\f{\sigma'(s)-\widetilde{\tau}(s)}{2}\pm\sqrt{\left(\f{\sigma'(s)-\widetilde{\tau}(s)}{2}\right)^2-\widetilde{\sigma}(s)+k\,\sigma(s)},\\
\tau(s)&=\widetilde{\tau}(s)+2\pi(s),~
\bar{\nu}=k+\pi'(s),
\ea 
\eeq
$\bar{\nu}$ and $k$ are real constants. Since $\pi(s)$ is a polynomial in $s$, we have to find $k$ in such a way that, $\left(\f{\sigma'(s)-\widetilde{\tau}(s)}{2}\right)^2-\widetilde{\sigma}(s)+k\,\sigma(s)$ is a square of a polynomial in $s$. Then the solutions of 
Eq.~(\ref{Eq.y}) are given by, 
\beq
\phi_{2,n_r}(s)=\f{1}{\rho(s)}\f{d^{n_r}}{ds^{n_r}}\left[\sigma^{n_r}(s)\rho(s)\right],
\eeq
and eigenvalues are obtained as, 
\beq\label{eigen.nu}
\nu_{n_r}=-{n_r}\,\tau'(s)-\f{{n_r}({n_r}-1)}{2}\sigma^{''}(s),~{n_r}=0,1,2,\dots, 
\eeq 
where
\beq
\ba{l}
\rho(s)=\ds \left[\sigma(s)\right]^{-1}\,e^{\int\f{\tau(s)}{\sigma(s)}ds}.
\ea 
\eeq  

\section{NU method for MR potential}\label{Sec.Examples1}
\subsection{Approximation 1: Greene-Aldrich-type}
Following \cite{qiang2007,chen2009,MR4.SMIkhdair,greene,wei2008,MR7.nc,MR8,qiang2009,MR10.qdot,MR11.thermal,MR12.scattering}, at first, let us consider, 
\beq
\ba{ll}\label{greene.app1}
\f{1}{r^2}&\approx f_1(r)=\f{1}{b^2}\left(x_{11}+\f{x_{21}e^{-r/b}}{1-e^{-r/b}}+\f{x_{31}e^{-2r/b}}{(1-e^{-r/b})^2}\right),
\ea 
\eeq
where
\beq
x_{11}= \f{1}{12},x_{21}=1,x_{31}=1,
\eeq 
when $ r/b\ll 1$. In the limiting case, $\lim\limits_{b\rightarrow \infty}\f{1}{b^2}\left(\f{1}{12}+\f{s}{1-s}+\f{s^2}{(1-s)^2}\right)=\f{1}{r^2}$. Therefore, for a fixed $b$, Eq.~(\ref{greene.app1}) is a good approximation near $r=0$. It is to be mentioned that, for $x_{11}=0,x_{21}=x_{31}=1$, this approximation corresponds to that of \cite{qiang2007,wei2008}, whereas for $x_{11}=0,x_{21}=e^{1/b},x_{31}=1$, it refers to \cite{MR8,qiang2009}. 

\subsection{Approximation 2: Pekeris-type}
In another attempt, $\f{1}{r^2}$ can be expressed \cite{Pekeris,MR3} around $r=r_0$, as, 
\beq\label{pekeris.app2}
\f{1}{r^2}\approx f_2(r)=\f{1}{b^2}\left(x_{12}+\f{x_{22}e^{-r/b}}{1-e^{-r/b}}+\f{x_{32}e^{-2r/b}}{(1-e^{-r/b})^2}\right),
\eeq 
where
\beq
\ba{ll}
x_{12}&=\ds\f{3-3u+u^2+\left(2u-6\right)s_0+\left(u+3\right)s_0^2}{u^4},\\
x_{22}&=\ds\f{2\left(1-s_0\right)^2}{u^4}\left(3+u+\f{2u-3}{s_0}\right),\\
x_{32}&=-\ds\f{\left(1-s_0\right)^3}{u^4}\left(\f{3+u}{s_0}+\f{u-3}{s_0^2}\right),\\
s_0&=e^{-u},~u=\f{r_0}{b}.
\ea
\eeq
In particular, if $r_0=b\ln\left[1+\f{2\a(\a-1)}{A}\right]$, then the potential value $V_{MR}(r_0)$ is minimum, which is equal 
to $-\f{A^2\hbar^2}{8\mu b^2\a(\a-1)}$ for $A>0$ and $\a\in(-\infty,0)\cup(1,\infty)$.

\subsection{Approximation 3: Pekeris-type}
In \cite{shd.pla.2008,MR4.SMIkhdair}, the authors considered an approximation of the following form,  
\beq\label{approx.app3}
\f{1}{r^2}\approx f_3(r)=\f{1}{b^2}\left(x_{13}+\f{x_{23}e^{-r/b}}{1-e^{-r/b}}+\f{x_{33}e^{-2r/b}}{\left(1-e^{-r/b}\right)^2}\right), 
\eeq
where
\beq
\ba{ll}
x_{13}&=\f{12\epsilon_1^2-4\epsilon_1(2A+3\epsilon_1)\log(\epsilon_2)+\epsilon_3^2\log(\epsilon_2)^2}{\epsilon_4^2\log(\epsilon_2)^4},\\
x_{23}&=\f{8\epsilon_1^2\left[-6\epsilon_1+(3A+4\epsilon_1)\log(\epsilon_2)\right]}{A\epsilon_4^2\log(\epsilon_2)^4},\\
x_{33}&=-\f{16\epsilon_1^3\left[-3\epsilon_1+\epsilon_3\log(\epsilon_2)\right]}{A^2\epsilon_4^2\log(\epsilon_2)^4},\\
\epsilon_1&=\a(\a-1),~\epsilon_2=1+\f{2\a(\a-1)}{A},\\
\epsilon_3&= A\epsilon_2,~
\epsilon_4=b\epsilon_3.
\ea
\eeq 

\subsection{Approximation 4: A linear combination of Greene-Aldrich and Pekeris-type}
\begin{figure}[htp] 
	\centering
	\includegraphics[width=18cm,height=16cm]{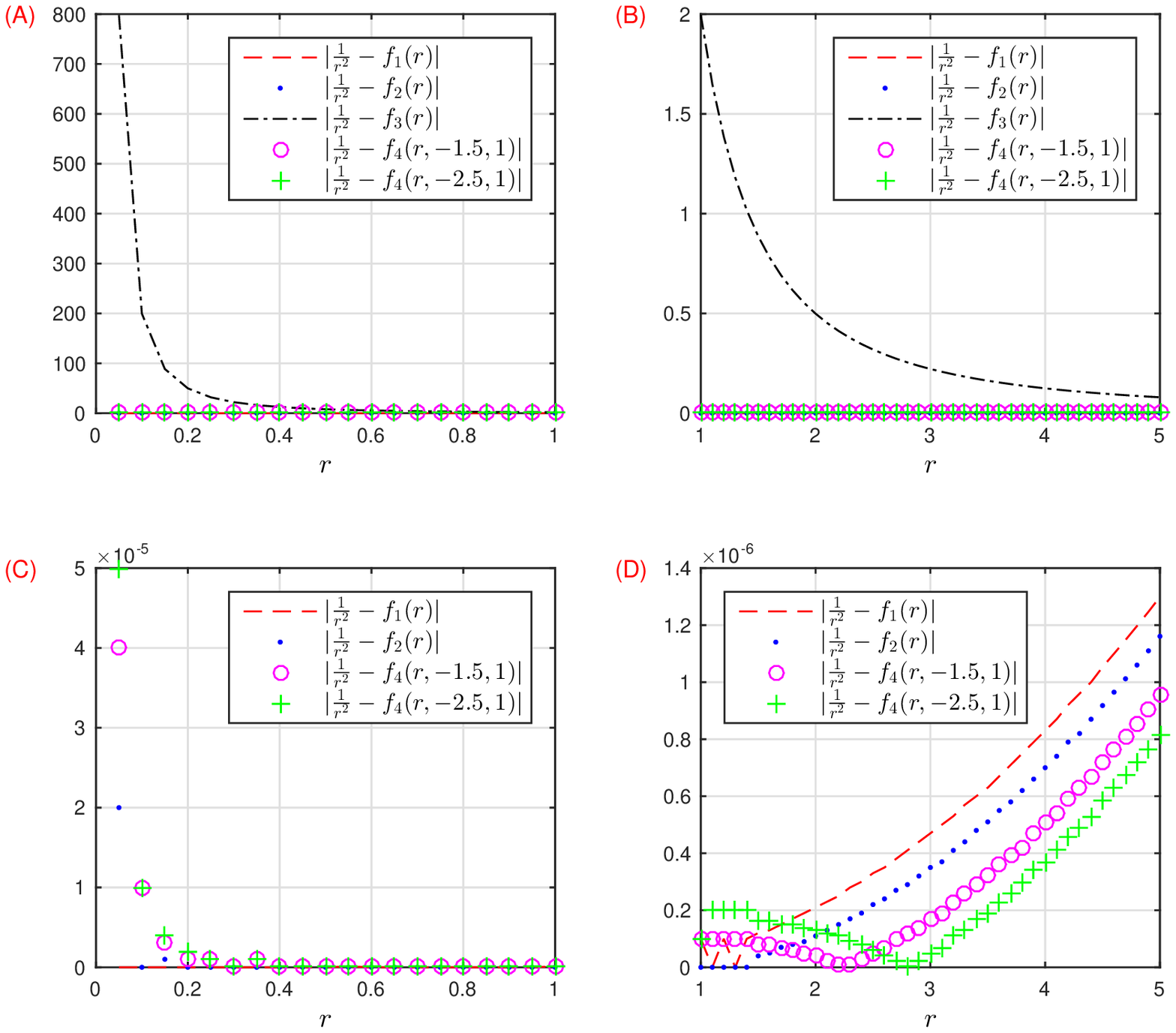}
	\caption{\label{Fig.1approximation.MR} Plot of the differences between exact centrifugal term $\f{l(l+1)}{r^2}$ and various approximations 
from Eqs.~(\ref{greene.app1}), (\ref{pekeris.app2}), (\ref{approx.app3}), (\ref{app.convex}) in MR potential, for $l=1,\hbar=\mu=1,\a=1.5,b=1/0.05, A=2b$, 
in panels (A)-(D) respectively.}
\end{figure}

Now, we propose a new approximation of the form \cite{nath2021}, 
\beq\label{app.convex}
\f{1}{r^2}\approx f_4(r,\lam,\nu)=\f{1}{b^2}\left(x_1+\f{x_2e^{-r/b}}{1-e^{-r/b}}+\f{x_3e^{-2r/b}}{(1-e^{-r/b})^2}\right),
\eeq 
where
\beq
\ba{l}
\left(\ba{l}x_1\\x_2\\x_3\ea\right)=\left(\ba{lll}x_{11} & x_{12}&x_{13}\\x_{21} & x_{22}&x_{23}\\x_{31} & x_{32}&x_{33}\ea \right)\left(\ba{l}\nu\lam\\\nu(1-\lam)\\1-\nu\ea\right),
\ea 
\eeq 
and $\lam,\nu$ are dimensionless real constants. In Fig.~\ref{Fig.1approximation.MR}, the centrifugal term under different 
approximations defined in Eqs.~(\ref{greene.app1}), (\ref{pekeris.app2}), (\ref{approx.app3}) and (\ref{app.convex}), is displayed. From this figure, 
it is clear that, Eq.~(\ref{app.convex}) gives the best approximation amongst all. If $(\lam,\nu)=(1,1)$, then Eq.~(\ref{app.convex}) 
implies Eq.~(\ref{greene.app1}); if $(\lam,\nu)=(0,1)$, then it leads to Eq.~(\ref{pekeris.app2}); if $(\lam,\nu)=(\lam,0)$, then it 
implies Eq.~(\ref{approx.app3}). The approximation is suitable for negative values of $\lam$ for large $r$. Furthermore, it is  
applicable for any values of $r$. 

\subsection{Solution of the MR potential}
Now, under the transformation, $s=e^{-r/b}$, Eq.(\ref{Eq.R}) becomes, 
\beq
\ba{l}\label{Eq.R.NU.MR}
\f{d^2R}{ds^2}+\f{1-s}{s(1-s)}\f{dR}{ds}+\f{-A_1s^2+Bs-C}{s^2(1-s)^2}R=0,
\ea 
\eeq 
where
\beq
\ba{ll}
A_1 &=\bar{\varepsilon}^2+A+\a(\a-1)+l(l+1)(x_3-x_2),\\
B& =2\bar{\varepsilon}^2+A-l(l+1) x_2,\\
C& =\bar{\varepsilon}^2=l(l+1) x_1-\varepsilon, 
\varepsilon=\f{2\mu b^2E}{\hbar^2}.
\ea
\eeq 
According to NU method, we obtain a pair of $k$, as given by \cite{ikhdair2008},
\beq
\ba{ll}\label{kpitaunu.MR}
k_{\pm}=A-l(l+1) x_2\pm\bar{\varepsilon}\sqrt{(2\a-1)^2+4l(l+1)x_3}.
\ea
\eeq
Applying Eq.~(\ref{kpitaunu.MR}) in Eq.~(\ref{pi.sigma}), we obtain,  
\beq
\ba{ll}\label{pi.tau.nu.MR}
\pi(s)
&=-\f{s}{2}\pm\left\{\ba{lll}\left(\bar{\varepsilon}-\f{1}{2}\sqrt{(2\a-1)^2+4l(l+1) x_3}\right)s+\bar{\varepsilon},&k=k_+,& k_+-B>0\\\left(\bar{\varepsilon}+\f{1}{2}\sqrt{(2\a-1)^2+4l(l+1) x_3}\right)s+\bar{\varepsilon},&k=k_-,& k_--B>0\\
\left(\bar{\varepsilon}-\f{1}{2}\sqrt{(2\a-1)^2+4l(l+1) x_3}\right)s-\bar{\varepsilon},&k=k_+,& k_+-B<0\\\left(\bar{\varepsilon}+\f{1}{2}\sqrt{(2\a-1)^2+4l(l+1) x_3}\right)s-\bar{\varepsilon},&k=k_-,& k_--B<0\ea \right\}.
\ea
\eeq
For the potential in Eq.~(\ref{pot.MR}), we have chosen $k=k_-$ with $k_--B<0$, and selected, 
\beq\label{rho.phi}
\pi(s)=-\f{s}{2}-\left(\bar{\varepsilon}+\f{1}{2}\sqrt{(2\a-1)^2+4l(l+1) x_3}\right)s+\bar{\varepsilon}.
\eeq
This gives,     
\beq
\ba{l}
\rho(s)=s^{2\bar{\varepsilon}}(1-s)^{2L-1},\\
\phi_1(s)=s^{\bar{\varepsilon}}(1-s)^L,
\ea 
\eeq
and
\beq
\ba{ll}
\phi_{2,n_r}&=({n_r})!\,P_{n_r}^{(2\bar{\varepsilon},2L-1)}(1-2s),\\
&=\f{\G\left(n_r+2\bar{\varepsilon}+1\right)}{\G\left(2\bar{\varepsilon}+1\right)}{}_2F_1\left(-n_r,n_r+2\bar{\varepsilon}+2L;2\bar{\varepsilon}+1;s\right),
\ea
\eeq 
where
\beq 
L=\f{1}{2}+\f{1}{2}\sqrt{(2\a-1)^2+4l(l+1) x_3}.
\eeq
The eigenvalues, $E_{n_r,l}^{MR}$, will then be generated from the following relation, 
\beq\label{deri.tau}
k+(2n_r+1)\pi'(s)=n_r^2.
\eeq
Finally, we obtain the ro-vibrational energy spectrum of MR potential, as a function of $(\hbar,\mu),(\a,b),(\lam,\nu)$, as given below, 
\beq\label{Energy.MR}
\ba{ll}
E_{n_r,l}^{MR}
&=x_5-\f{\hbar^2}{2\mu b^2}\left(\f{x_4^2}{(n_r+L)^2}+\f{(n_r+L)^2}{4}\right),
\ea 
\eeq
where
\beq
\ba{ll}
x_4=\f{1}{2}\left[A+\a(\a-1)+l(l+1)(x_3-x_2)\right],x_5=\f{\hbar^2}{2\mu b^2}(l(l+1) x_1+x_4).
\ea
\eeq  
Therefore, the radial wave function of MR potential can be expressed as \cite{Gradshteyn}, 
\beq
R_{n_r}^{MR}(r)=N_{{n_r},l}\,s^{\bar{\varepsilon}}(1-s)^{L}\,{}_2F_1\left(-{n_r},{n_r}+2\bar{\varepsilon}+2L;2\bar{\varepsilon}+1;s\right),
\eeq 
where 
\beq
N_{{n_r},l}=\left[\f{2\bar{\varepsilon}(n_r+\bar{\varepsilon}+L)\G(n_r+2\bar{\varepsilon}+1)\G(n_r+2\bar{\varepsilon}+2L)}{({n_r})!b(n_r+\bar{\varepsilon})\G(n_r+2L)\left[\G(2\bar{\varepsilon}+1)\right]^2}\right]^{\f{1}{2}},
\eeq 
is the normalization constant, to be obtained from, 
\beq
\ds\int|\psi_{n_r,l,m}^{MR}({\bf r})|^2\,d{\bf r}=1,~ d{\bf r}=r^2\sin\theta\,dr\,d\theta\,d\phi. 
\eeq 
Finally, the explicit form of eigenfunctions of MR potential can be written as, 
\beq
\psi_{n_r,l,m}^{MR}({\bf r})=\left[\f{(2l+1)(l-|m|)!}{4\pi(l+|m|)!}\right]^{\f{1}{2}}\f{N_{{n_r},l}}{r}\,s^{\bar{\varepsilon}}(1-s)^{L}\,{}_2F_1\left(-{n_r},{n_r}+2\bar{\varepsilon}+2L;2\bar{\varepsilon}+1;s\right)\,P_{l}^m(\cos\theta)\,e^{im\phi},~s=e^{-r/b}.
\eeq
It is to be noted that the approximation (\ref{app.convex}) is well defined {\bf for $\lam, \nu$ satisfying,}
\beq
\ba{l}\label{relation1}
(2\a-1)^2+4l(l+1)\left(x_{31}\nu\lam+x_{32}\nu(1-\lam)+x_{33}(1-\nu)\right)>0,\\
x_{11}\nu\lam+x_{12}\nu(1-\lam)+x_{13}(1-\nu)>0.
\ea
\eeq 
\section{NU method for PT potential}\label{Sec.Examples2}
\subsection{Approximation 1: Greene-Aldrich type} 
At first, we consider an approximation \cite{shd.ijmpa2008} of the form, 
\beq\label{ap1}
\f{1}{r^2}\approx
f_1(r)=\a^2\left(x_{11}+\f{x_{21}}{\sinh^2\a r}+\f{x_{31}\cosh\a r}{\sinh^2\a r}\right),
\eeq 
where
\beq
x_{11}=0,x_{21}=\f{1}{2},x_{31}=\f{1}{2}.
\eeq

\subsection{Approximation 2: Greene-Aldrich type}
Next let us examine another Greene-type approximation due to \cite{greene,PT2.epjp2020}, 
\beq\label{ap2}
\f{1}{r^2}\approx f_2(r)= 
\a^2\left(x_{12}+\f{x_{22}}{\sinh^2\a r}+\f{x_{32}\cosh\a r}{\sinh^2\a r}\right),
\eeq 
where
\beq
x_{12}=\f{1}{12},x_{22}=\f{1}{2},x_{32}=\f{1}{2}.
\eeq

\subsection{Approximation 3: Pekeris-type} 
The Pekeris-type approximation, around $r=r_0$ is given by \cite{PT2.epjp2020,Pekeris,pekeris.badawi,pekeris.ap2,pekeris.ap3,pekeris.ap4}, 
\beq\label{ap3}
\f{1}{r^2}\approx f_3(r)=
\a^2\left(x_{13}+\f{x_{23}}{\sinh^2\a r}+\f{x_{33}\cosh\a r}{\sinh^2\a r}\right),
\eeq 
where
\beq
\ba{ll}
x_{13}=\ds\f{1}{u^4}\left(3+u^2-3u\coth u\right),\\
x_{23}=\ds\f{1}{4u^4}\left(18+6\cosh(2u)-23u\coth u-u\cosh(3u) \ \mathrm{cosech} \ u\right),\\
x_{33}=\ds\f{1}{u^4}\left(4u+2u\cosh(2u)-3\sinh(2u)\right) \ \mathrm{cosech} \ u,\\
u=\a r_0.
\ea
\eeq
In particular, if $r_0=\f{1}{\a}\tanh^{-1}\left(\sqrt{\f{2\sqrt{\xi_1^2-\xi_2^2}}{\xi_1+\sqrt{\xi_1^2-\xi_2^2}}}\right)$, then the potential has a 
minimum at $r=r_0$.

\subsection{Approximation 4: A linear combination of Greene-Aldrich and Pekeris-type}
\begin{figure}[htp] 
	\centering
	\includegraphics[width=18cm,height=10cm]{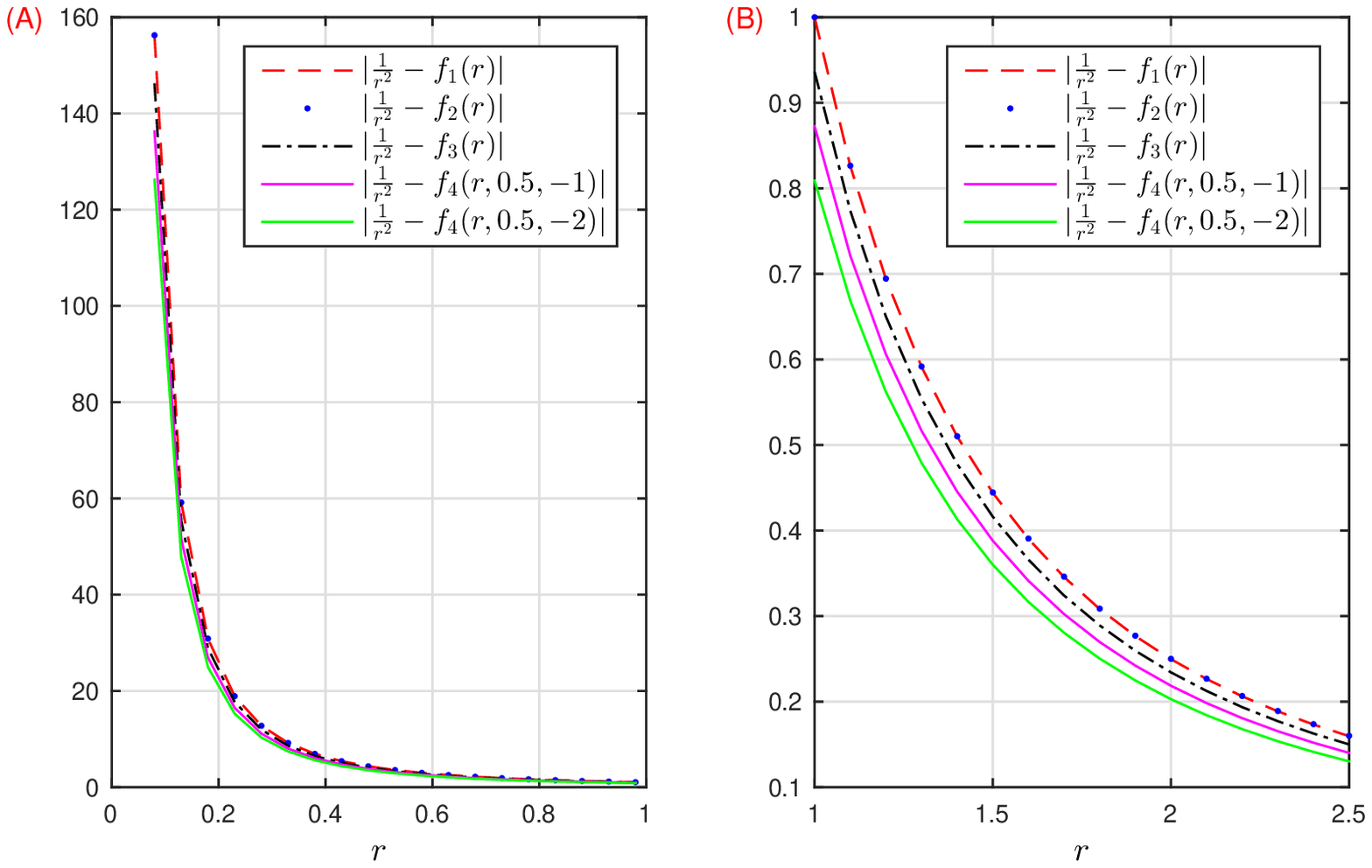}
	\caption{\label{Fig.2approximation.PT} Plot of the differences between exact centrifugal term $\f{l(l+1)}{r^2}$ and various approximations 
from Eqs.~(\ref{ap1}), (\ref{ap2}), (\ref{ap3}) and (\ref{ap4}) in PT potential, for $l=1,\hbar=\mu=1,\xi_1=4,\xi_2=2,\a=0.05$, in 
panels (A)-(D) respectively.}
\end{figure}

Now let us discuss a linear combination of three approximations in Eqs.~(\ref{ap1}), (\ref{ap2}), (\ref{ap3}), proposed lately \cite{nath2021},  
\beq\label{ap4}
\f{1}{r^2}\approx f_4(r,\lam,\nu)=
\a^2\left(x_{1}+\f{x_{2}}{\sinh^2\a r}+\f{x_{3}\cosh\a r}{\sinh^2\a r}\right),
\eeq 
where
\beq
\left(\ba{c}x_1\\x_2\\x_3\ea\right)=\left(\ba{ccc}x_{11} & x_{12} & x_{13}\\
x_{21} & x_{22} & x_{23}\\
x_{31} & x_{32} & x_{33}\ea\right)\left(\ba{l}\nu\lam\\\nu(1-\lam)\\1-\nu\ea\right),
\eeq 
and $\lam,\nu$ are real constants. In Fig.~\ref{Fig.2approximation.PT}, the centrifugal term under different approximations, defined in 
Eqs.~(\ref{ap1}), (\ref{ap2}), 
(\ref{ap3}) and (\ref{ap4}) are displayed. It is quite clear that, the approximation of Eq.~(\ref{ap4}) is the best of all. When $(\lam,\nu)=(1,1)$, 
Eq.~(\ref{ap4}) implies Eq.~(\ref{ap1}); $(\lam,\nu)=(0,1)$ means (\ref{ap2}); $(\lam,\nu)=(\lam,0)$, leads to (\ref{ap3}). It may be noted that, 
Eq.~(\ref{ap4}) is good for negative $\nu$ in the large-$r$ region. The advantage is that, for any arbitrary values of $r$, this approximation is 
easily seen to provide the best representation amongst all, for select values of $\lam$ and $\nu$. 

\subsection{Solution of the PT potential}
Under the transformation, $s=\tanh^2\f{\a r}{2}$, the radial Schr\"odinger equation in (\ref{Eq.R}) becomes,  
\beq
\ba{l}\label{Eq.R.NU.PT}
\f{d^2R}{ds^2}+\f{1-3s}{s(1-s)}\f{dR}{ds}+\f{-As^2+Bs-C}{s^2(1-s)^2}R=0,
\ea 
\eeq 
where
\beq
\ba{ll}
A &=\f{\mu}{2\a^2\hbar^2}(\xi_1+\xi_2)+\f{l(l+1)}{4}(x_3-x_2),\\
B &=\f{2\mu}{\a^2\hbar^2}\left(E+\f{\xi_1}{2}\right)+\f{l(l+1)}{2}(x_2-2x_1),\\
C &=\f{\mu}{2\a^2\hbar^2}(\xi_1-\xi_2)+\f{l(l+1)}{4}(x_3+x_2). \\
\ea
\eeq 
Similarly, we obtain a pair of $k$, which can be defined by, 
\beq\label{kpitaunu}
k_{\pm}=B-2C\pm a\sqrt{C},~a=\sqrt{1+4(A-B+C)}.
\eeq
This produces, 
\beq
\ba{ll}\label{pi.tau.nu.PT}
\pi(s)&=\f{s}{2}\pm\left\{\ba{lll}\left|\sqrt{C}-\f{a}{2}\right|s+\sqrt{C},&k=k_+,& k_+-B>0\\\left(\sqrt{C}+\f{a}{2}\right)s+\sqrt{C},&k=k_-,& k_--B>0\\
\left|\sqrt{C}-\f{a}{2}\right|s-\sqrt{C},&k=k_+,& k_+-B<0\\\left(\sqrt{C}+\f{a}{2}\right)s-\sqrt{C},&k=k_-,& k_--B<0\ea \right\}.
\ea
\eeq
Since $\tau'(s)<0$, for PT potential, we choose $k=k_-=B-2C-a\sqrt{C}$, where $k_--B<0$, and select $\pi(s)$ as, 
\beq
\ba{ll}
\pi(s)&=\left(\f{1-a}{2}-\sqrt{C}\right)s+\sqrt{C}.
\ea
\eeq  
Then, we find, 
\beq
\ba{l}\label{rho.phi.PT}
\tau(s)=1+2\sqrt{C}-\left(2+a+2\sqrt{C}\right)s,\\
\rho(s)=s^{2\sqrt{C}}(1-s)^{a},\\
\phi_1(s)=s^{\sqrt{C}}(1-s)^{\f{a-1}{2}},
\ea 
\eeq
along with
\beq
\ba{ll}
\phi_{2,n_r}(s)&=s^{-2\sqrt{C}}(1-s)^{-a}\f{d^{n_r}}{ds^{n_r}}\left[s^{{n_r}+2\sqrt{C}}(1-s)^{{n_r}+a}\right]=({n_r})!\,P_{n_r}^{(2\sqrt{C},a)}(1-2s).
\ea
\eeq 
Then the ro-vibrational energy $E_{n_r,l}^{PT}$ of PT potential can be obtained from the following relation, 
\beq\label{deri.tau.PT}
k+(2{n_r}+1)\pi'(s)={n_r}^2+2{n_r}.
\eeq
Accordingly, one gets the ro-vibrational energy spectrum, as function of $(\hbar,\mu),(\xi_1,\xi_2,\a),(\lam,\nu),(n_r,l)$, as below, 
\beq\label{Energy.PT}
E_{n_r,l}^{PT}=x_4-\f{\a^2\hbar^2}{2\mu}(n_r+L)^2,
\eeq 
where
\beq
\ba{ll}
x_4&=-\f{\xi_1}{2}+\f{l(l+1)\a^2\hbar^2}{4\mu}\left(2x_1-x_2\right)+\f{\a^2\hbar^2}{2\mu}\left[\f{1}{4}+A+C\right],\\
L&=\f{1}{2}+\sqrt{C}-\sqrt{1+A}.
\ea
\eeq  
Therefore, the explicit form of radial wave function can be found to be \cite{Gradshteyn}, 
\beq
R_{n_r}^{PT}(r)=N_{n_r,l}\,s^{\sqrt{C}}(1-s)^{\f{a-1}{2}}P_{n_r}^{(2\sqrt{C},a)}(1-2s),
\eeq 
where 
\beq
\ba{r}
N_{n_r,l}= \sqrt{\a}\Big(\ds\sum\limits_{m=0}^{n_r}(-1)^{n_r-m}\binom{n_r+\g}{m}\binom{n_r+\d}{n_r-m}\f{\G(n_r-m+\g+1)\G(m+\d+1)\G(n_r+\g+\f{3}{2})}{n_r!\G(n_r+\g+\d+2)\G(n_r-m+\g+1)}\\
 \ds\times {}_3F_2(-n_r,2n_r+1+\g+\d,m+1+\d;n_r-m+\g+1,n_r+\g+\d+2;1)\Big)^{-\f{1}{2}}
\ea 
\eeq 
is the normalization constant determined from the following condition,  
\beq
\ds\int \left[R_{n_r}^{PT}(r)\right]^2\,dr=1, 
\eeq 
where
\beq
\g=2\sqrt{C}-\f{1}{2},~\d=a-2.
\eeq
The required eigenfunctions of PT potential are finally expressed as, 
\beq
\ds\psi_{n_r,l,m}^{PT}({\bf r})=\left[\f{(2l+1)(l-|m|)!}{4\pi(l+|m|)!}\right]^{\f{1}{2}}\f{N_{n_r,l}}{r}\,s^{\sqrt{C}}(1-s)^{\f{a-1}{2}}P_{n_r}^{(2\sqrt{C},a)}(1-2s)\,P_{l}^m(\cos\theta)\,e^{im\phi},~s=\tanh^2\left(\f{\a r}{2}\right). 
\eeq
It is to be noted that the approximation (\ref{ap4}) is well defined {\bf for $\lam, \nu$ satisfying,}
\beq
\ba{l}\label{relation2}
2\mu(\xi_1-\xi_2)+\a^2\hbar^2l(l+1)\left[(x_{21}+x_{31})\nu\lam+(x_{22}+x_{32})\nu(1-\lam)+(x_{23}+x_{33})(1-\nu)\right]>0,\\
(2x_{11}-x_{21}+x_{31})\nu\lam+(2x_{12}-x_{22}+x_{32})\nu(1-\lam)+(2x_{13}-x_{23}+x_{33})(1-\nu)>0
\ea 
\eeq
\section{Results and discussion}\label{Sec.Results}
In Table~\ref{Table1}, computed energies, $E_{n_r,l}^{MR}$ of the MR potential are given. 
The performance of the proposed schemes are illustrated for eight representative states for five different sets of ($\lam$, $\nu$) pair; 
namely, (1,1), (0,1), (1,0), ($-$1.5,1), ($-$2.5,1), which include the negative approximation parameters as well. It appears that for all the 
states under consideration, the (1,0) set somehow separates out from all others, which remain in a family of their own. The energies corresponding 
of (1,1) and (0,1) sets in columns 3 and 4 match quite well with references \cite{MR8} and \cite{MR3}, which is duly pointed out in footnote. {\bf Additional reference energies are also provided from the numerical works of Lucha \cite{lucha1999} and generalized pseudospectral method 
(GPS) \cite{MR9}}. Similar energies are presented for 
PT potential in Table~\ref{Table2}, again for same eight states of previous table. In addition to the first three positive $(\lam, \nu)$ parameter 
sets of Table~\ref{Table1}, in this case, we consider negative sets as (0.5, $-$1) and (0.5, $-$2). Once again, the energies of column 3 having (1,1) parameter set 
compares well those from \cite{shd.ijmpa2008}, as indicated in footnote. {\bf These energies are also compared with the accurate results from GPS method, which has been very successful for a number of model and real systems \cite{MR9,gps,gps2}}. One finds that the current proposed approximation in 
Eq.~(\ref{app.convex}) for MR potential fares better for negative $\lam$ whereas the same for PT potential in Eq.~(\ref{ap4}) works better 
for negative $\nu$.   

\begin{figure}[htp] 
	\centering
	\includegraphics[width=10cm,height=10cm]{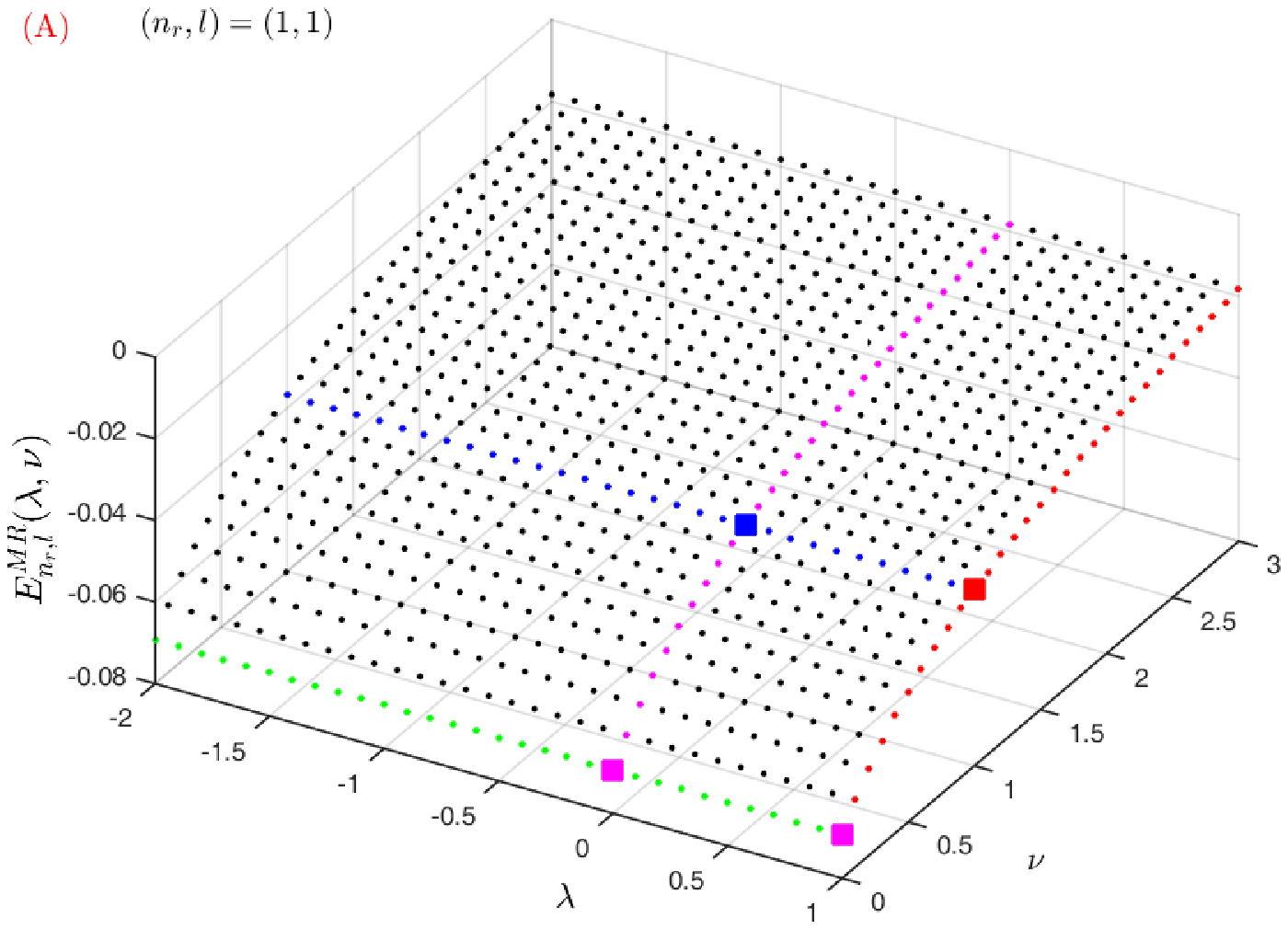}~\includegraphics[width=10cm,height=10cm]{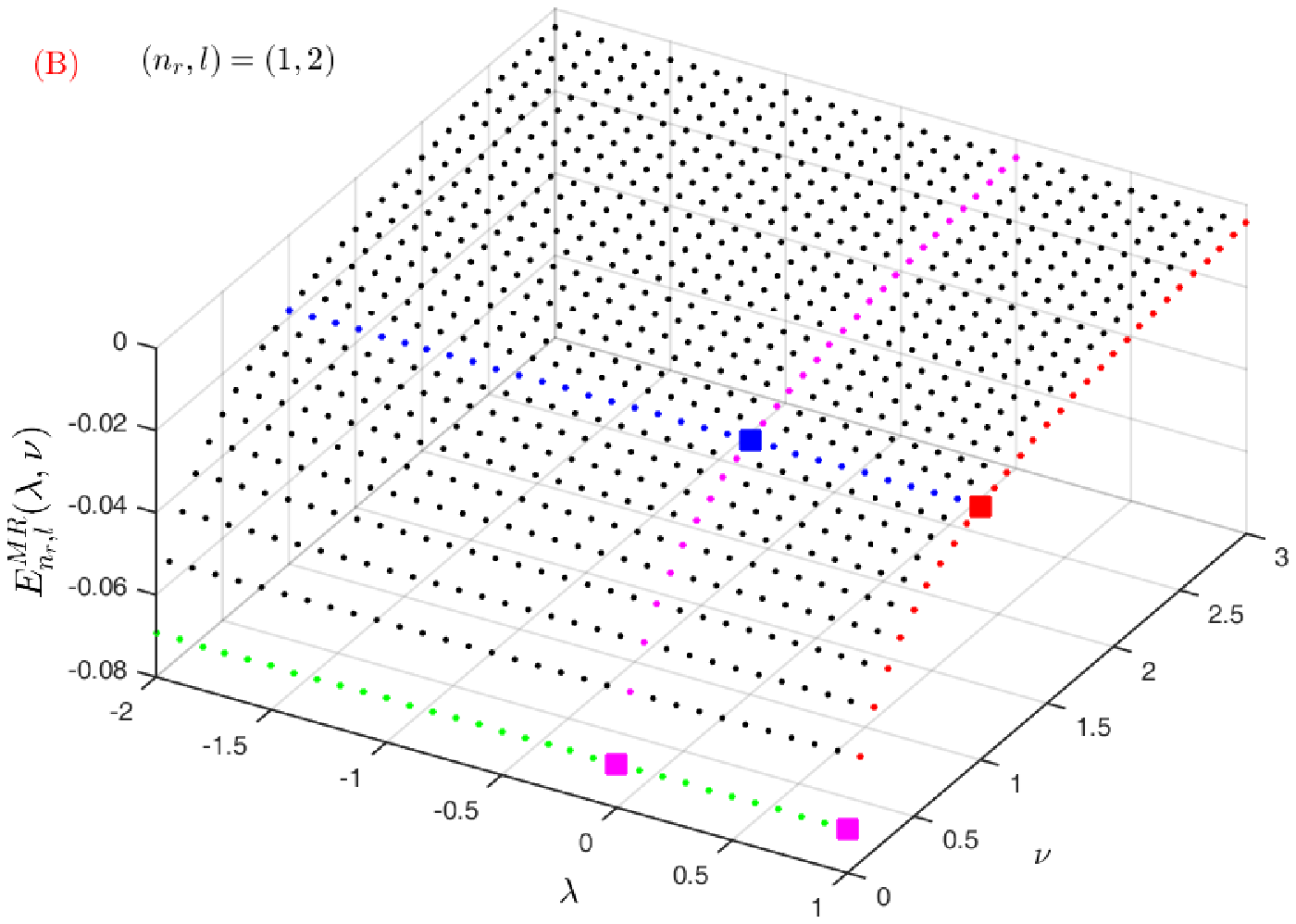}\\
	\includegraphics[width=10cm,height=10cm]{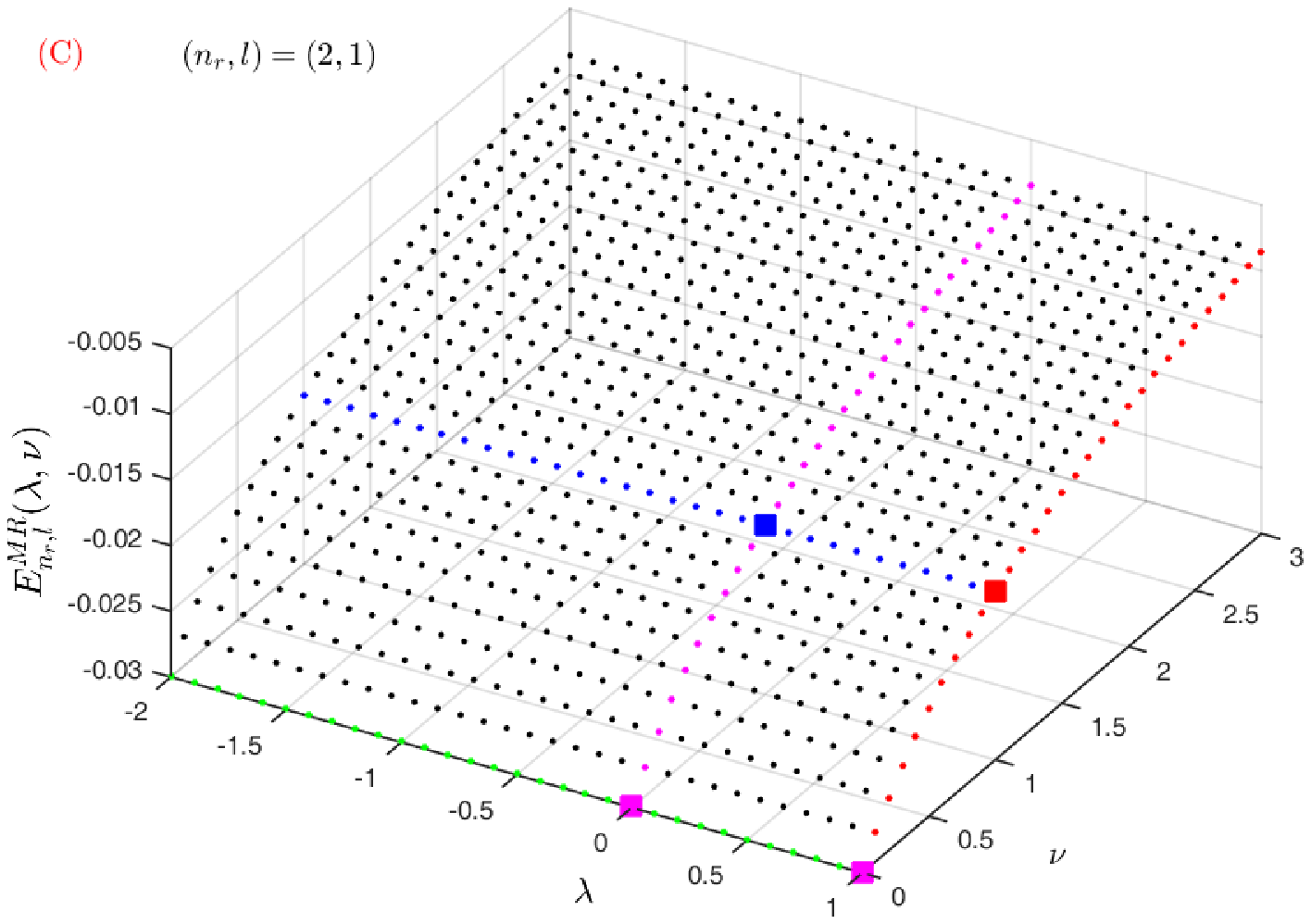}~\includegraphics[width=10cm,height=10cm]{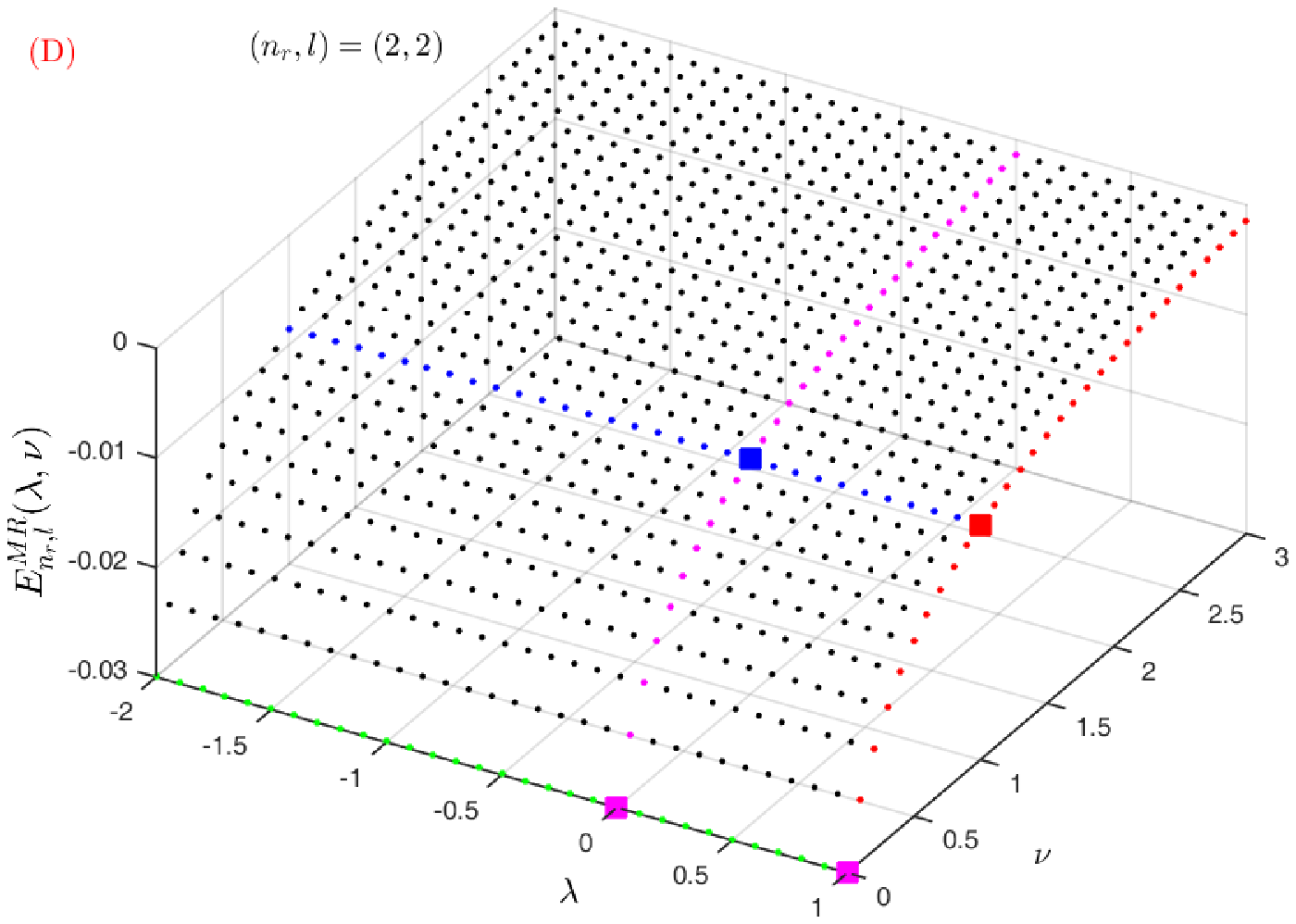}
	\caption{\label{Fig.3MRP.Energy} Effect of $\lam$ and $\nu$ on the energy, $E_{n_r,l}^{MR}$, of MR potential. Panels 
(A), (B), (C), (D) correspond to $(n_r, l)$ values (1,1), (1,2), (2,1) and (2,2) respectively. The parameters are: 
$\hbar=\mu=1,\a=1.5,b=1/0.025$. See text for details.}
\end{figure}

\begin{figure}[htp] 
	\centering
	\includegraphics[width=10cm,height=10cm]{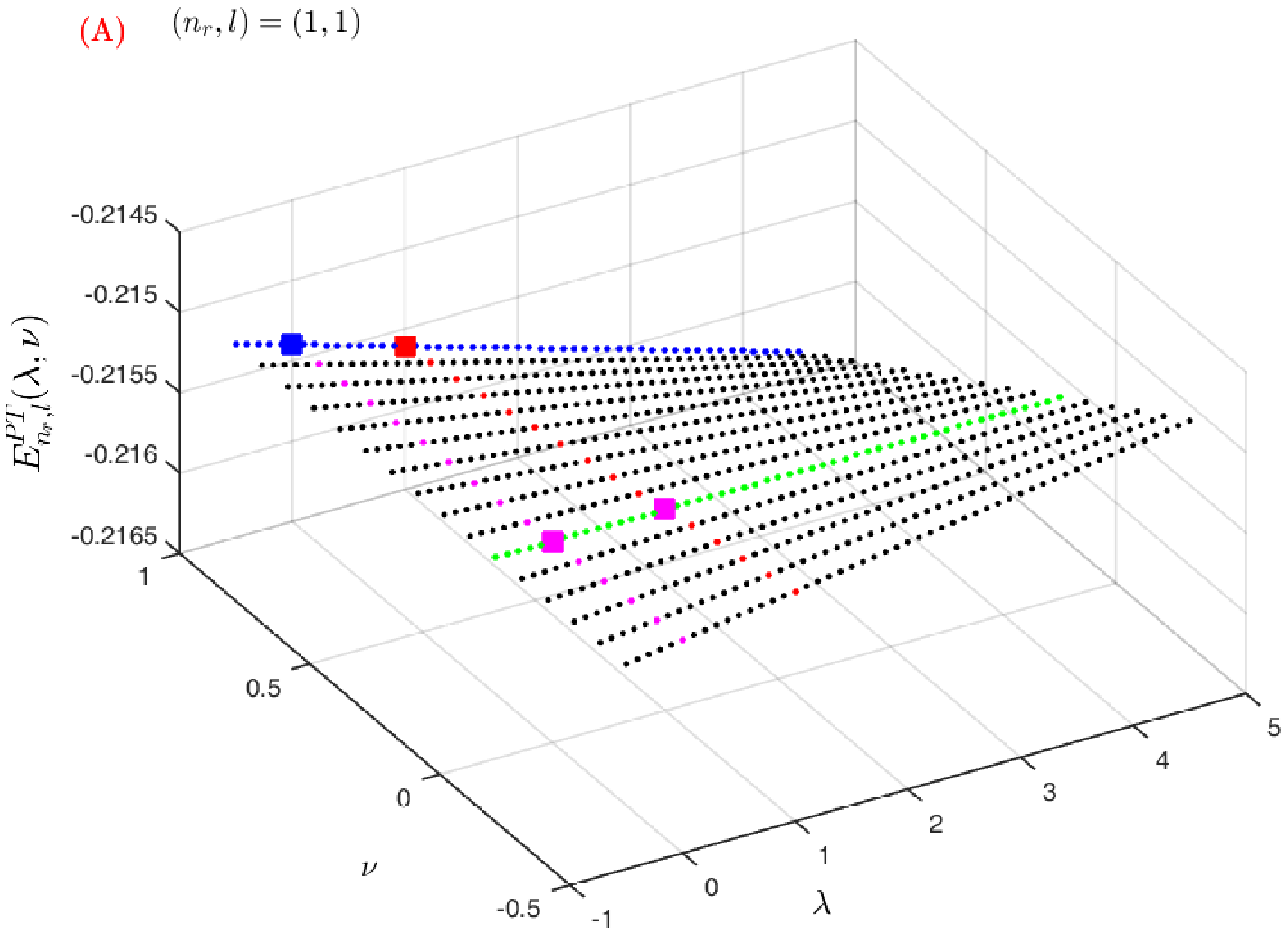}~\includegraphics[width=10cm,height=10cm]{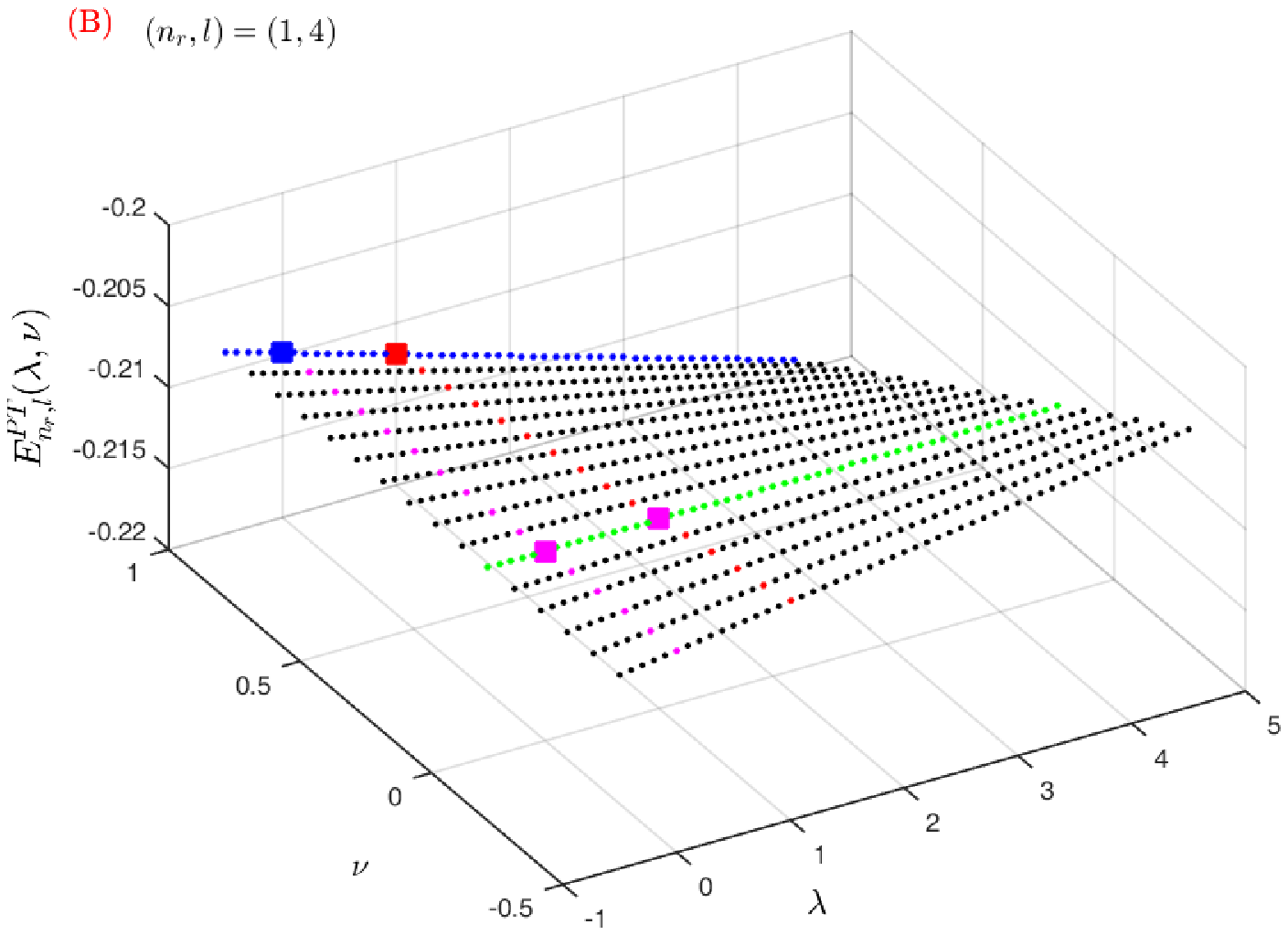}\\
	\includegraphics[width=10cm,height=10cm]{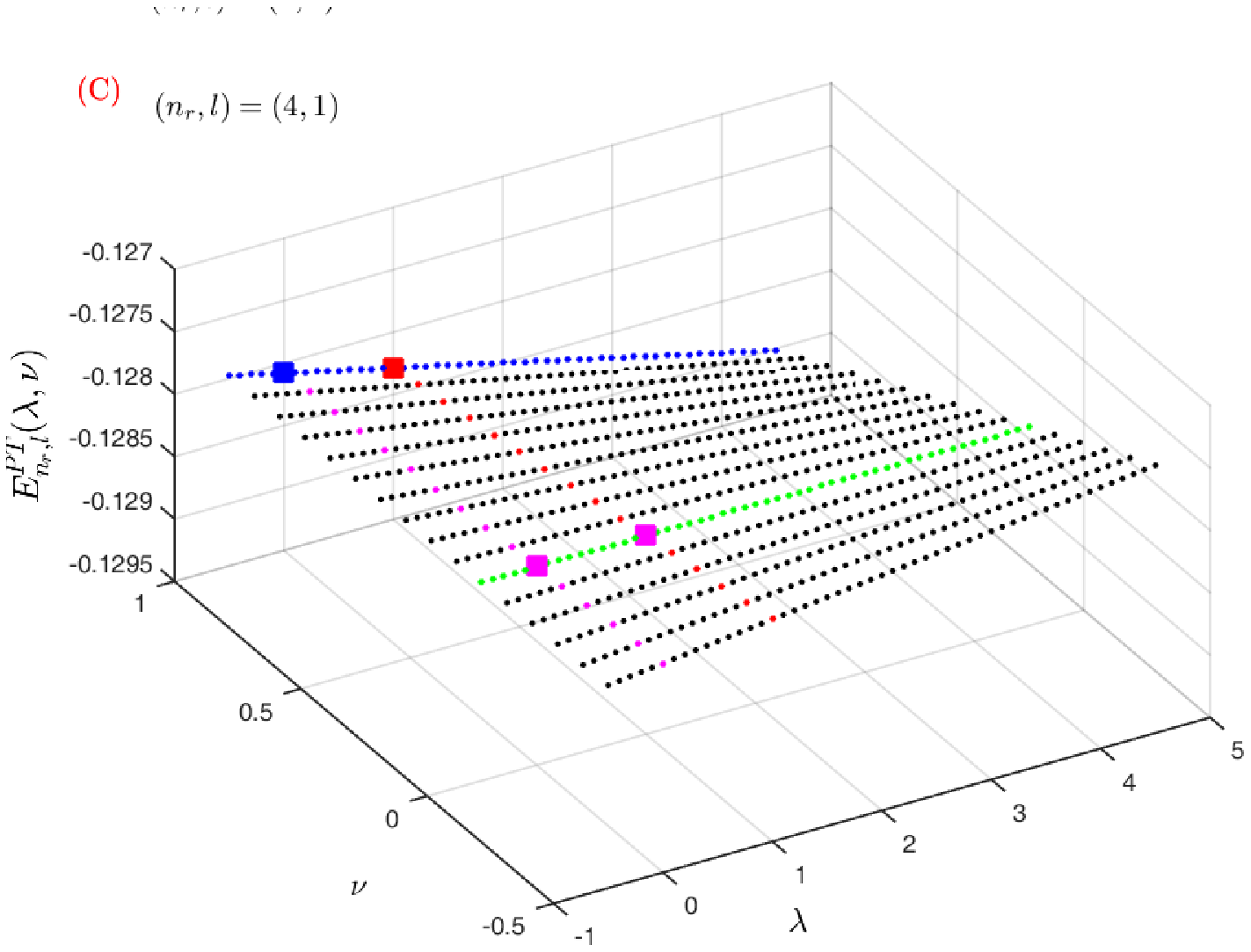}~\includegraphics[width=10cm,height=10cm]{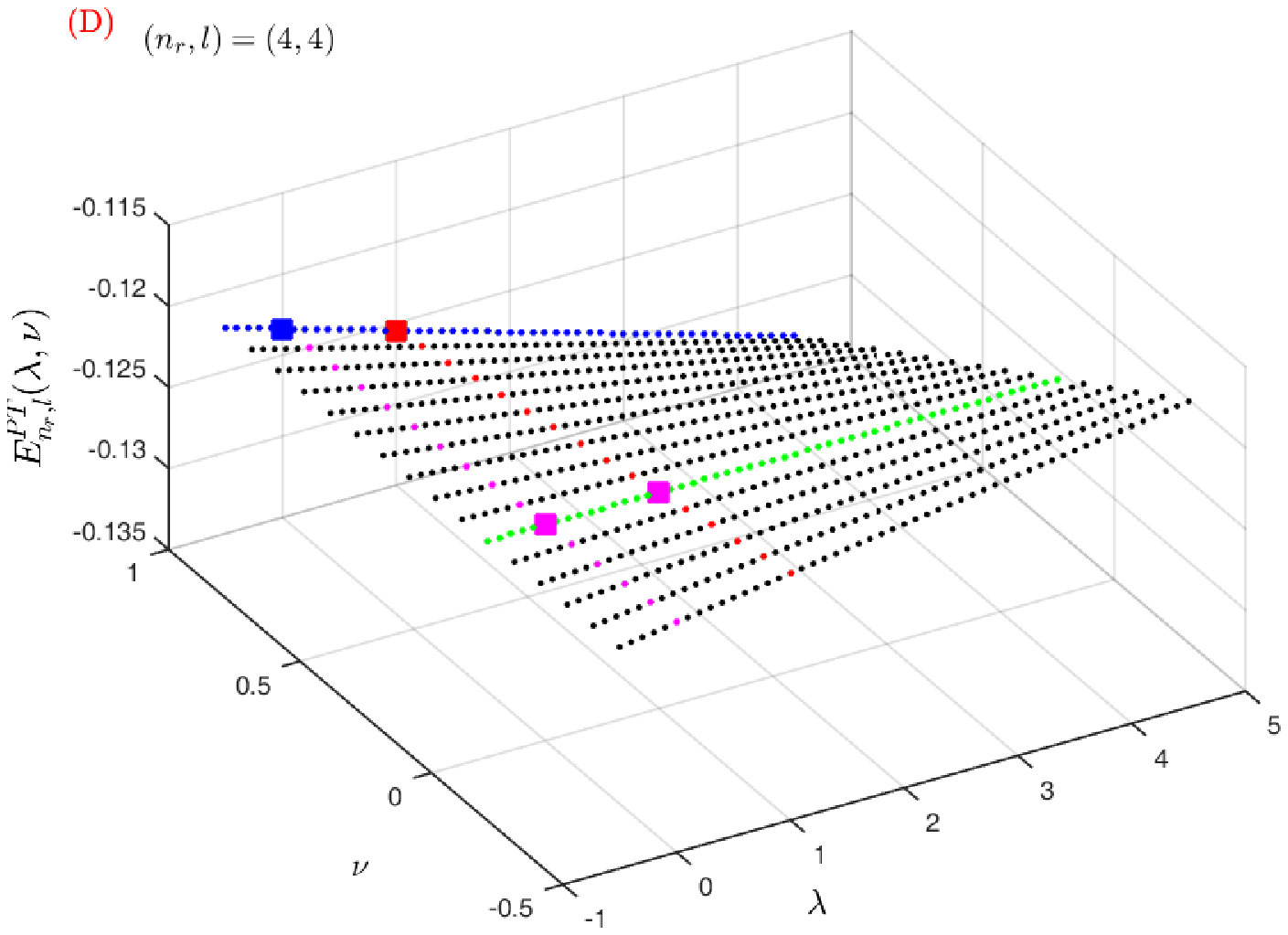}
	\caption{\label{Fig.4PTP.Energy} Effect of $\lam$ and $\nu$ on the energy, $E_{n_r,l}^{PT}$, of PT potential. Panels 
(A), (B), (C), (D) correspond to $(n_r, l)$ values (1,1), (1,4), (4,1) and (4,4) respectively. The parameters are: 
$\hbar=\mu=1,\xi_1=4,\xi_2=2,\a=0.05$. See text for details.}
\end{figure}

In order to examine the effects of $\lam$ and $\nu$, in Fig.~\ref{Fig.3MRP.Energy}, we have plotted the computed energies of MR potential
with respect to these two parameters. Four states corresponding to $(n_r, l)$ quantum numbers as (1,1), (1,2), (2,1) and (2,2) are displayed. 
One sees that, for a given $\lam$, $E_{n_r,l}^{MR}$ is an increasing function of $\nu$, and $E_{n_r,l}^{MR}\rightarrow 0$ as 
$\nu$ assumes larger values. Note that negative $\lam$ values are also considered. The energies marked with red, blue and magenta squares 
refer to $(\lam,\nu)=(1,1)$, (0,1) and \{(0,0), (1,0)\} respectively. These are in good harmony with the values presented in references 
\cite{MR8} and \cite{MR3}. Analogous plots are offered for PT potential in Fig.~\ref{Fig.4PTP.Energy}. In this 
occasion, the energy increases as $\lam$ increases for a specific $\nu<0$; while it decreases for a fixed $\nu>0$ as $\lam$ increases.  
The red, green and magenta squares in the diagram correspond to same $(\lam, \nu)$ pairs of previous figure. They recover the energies 
of \cite{shd.ijmpa2008} well. Also these energies are found to be in good agreement with \cite{pekeris.ap2}, for the parameter sets provided therein.  

\section{Conclusions}\label{Sec.Conclusion}
In this article, we have introduced a new simple novel approximations to the centrifugal term for both MR and PT potentials. These are intuitively derived from a 
linear combination of the commonly used Greene-Aldrich and Pekeris-type approximations. From this, the original approximations are recovered 
for certain special values of the two approximating parameters $\lam$ and $\nu$. Approximate analytical expressions are then presented for these two 
potentials by the NU method. It is gratifying to note that, the approximation perform quite nicely throughout the whole range of $r$, 
whereas, Greene-Aldrich and Pekeris provide superior approximations near the origin $r=0$, and $r=r_0$ (where the potential is minimum) respectively. Analytical expressions are
presented for eigenvalues and eigenfunctions.  

An investigation of the controlling parameters, $\lam$ and $\nu$ on energy spectra shows that, $E_{n_r,l}^{MR}$ is an increasing function of $\nu$ subject to the conditions (\ref{relation1}).
Whereas $E_{n_r,l}^{PT}$ is a increasing function of $\lam$ for $\nu<0$ and a decreasing function of $\lam$ for $\nu>0$ subject to the conditions (\ref{relation2}). 
For some special cases $(\lam,\nu)=(1,1)$, $(\lam,\nu)=(0,1)$, and $(\lam,\nu)=(\lam,0)$, energies of MR and PT potentials compare quite favorably 
with available literature results. It may be worthwhile to study the performance and efficacy of this approach for other related potentials of 
physical and chemical interest. Also its relevance in the thermodynamic studies may be pursued.   

\section*{Acknowledgement}
AKR gratefully acknowledges financial support from MATRICS, DST-SERB, New Delhi (sanction order: MTR/2019/000012). We thank the 
anonymous referee for constructive comments and suggestions. \\

\begin{table}[htp]
	\caption{\label{Table1} Energies, $E_{n_r,l}^{MR}$, of MR potential for five sets of $(\lam, \nu)$, given in parentheses. Here 
$\hbar=\mu=1,\a=1.5,b=1/0.025$. }
	\centering
	\begin{tabular}{|lr|lllllll|}\hline 
		$n_r$	& $l$ & $-E_{n_r,l}^{MR}(1,1)$\footnotemark[1] & $-E_{n_r,l}^{MR}(0,1)$\footnotemark[2] & $-E_{n_r,l}^{MR}(1,0)$ & $-E_{n_r,l}^{MR}(-1.5,1)$ & $-E_{n_r,l}^{MR}(-2.5,1)$& GPS \cite{MR9}   & Numerical \cite{MR4.SMIkhdair}\footnotemark[3] \\   \hline
	1  &1  & 0.036913014 &  0.036913019  & 0.069378105 &  0.036913027  & 0.036913032&0.036913922&    0.0369134           \\
	
	1  &2  & 0.018208662 &  0.018208677  & 0.069378064 &  0.0182087  & 0.018208715&0.0182117637&  0.0182115             \\
	
	1  &3  & 0.0086497057 &  0.0086497369  & 0.069378003 &  0.0086497836  & 0.0086498148&0.0086619417&  0.0086619              \\
	
	1  &4  & 0.003521352 &  0.0035214045  & 0.069377922 &  0.0035214833  & 0.0035215358&0.0035623305&   0.0035623        \\
	
	2  &1  & 0.017172833 &  0.017172838  & 0.029925043 &  0.017172846  & 0.017172851&0.0171740303&  0.0171740         \\
	
	2  &2  & 0.0085339943 &  0.0085340099  & 0.029925003 &  0.0085340333  & 0.0085340489&0.0085414805& 0.0085415       \\
	
	2  &3  & 0.0036481309 &  0.0036481624  & 0.029924942 &  0.0036482097  & 0.0036482412&0.0036774476& 0.0036774        \\
	
	2  &4  & 0.00093625137 &  0.00093630425  & 0.02992486 &  0.00093638357  & 0.00093643645&0.0010296092&   ---   \\    \hline	
	\end{tabular}
\begin{tabbing}
	\footnotemark[1]{These energies nearly coincide with those from \cite{MR8}}. \ \ \ \ \ \ \ \ \ \ \ \ \ \ \ \ \=
        \footnotemark[2]{These energies compare with those from \cite{MR3}.} \\ 
        \footnotemark[3]{These correspond to numerical results using the method of \cite{lucha1999}.} \\ 
\end{tabbing} 
\end{table}	

\begin{table}[htp]
\caption{\label{Table2} Energies, $E_{n_r,l}^{PT}$, of PT potential for five sets of $(\lam, \nu)$, given in parentheses. Here,
$\hbar=\mu=1,\xi_1=4,\xi_2=2,\a=0.05$.}
\centering
\begin{tabular}{|lr|lllllll|}\hline
$n_r$ & $l$ & $E_{n_r,l}^{PT}(1,1)$\footnotemark[1] & $E_{n_r,l}^{PT}(0,1)$ & $E_{n_r,l}^{PT}(1,0)$ & $E_{n_r,l}^{PT}(0.5,-1)$ & $E_{n_r,l}^{PT}(0.5,-2)$ &   GPS\footnotemark[2]  &     Lucha/others    \\            \hline
1  &1  & $-$0.21560894 &  $-$0.21540061  & $-$0.21524815 &  $-$0.21499153  & $-$0.21473492&0.215258812&\\

1  &2  & $-$0.21478931 &  $-$0.21416431  & $-$0.2137065 &  $-$0.21293627  & $-$0.21216612& 0.2141062447&\\

1  &3  & $-$0.2135647 &  $-$0.2123147  & $-$0.21139777 &  $-$0.20985618  & $-$0.20831492& 0.212382835&\\

1  &4  & $-$0.21194083 &  $-$0.2098575  & $-$0.20832642 &  $-$0.2057546  & $-$0.20318371& 0.2100950625&\\

2  &1  & $-$0.18400157 &  $-$0.18379323  & $-$0.18363528 &  $-$0.18337317  & $-$0.18311107&0.1836932855&\\

2  &2  & $-$0.18324436 &  $-$0.18261936  & $-$0.18214507 &  $-$0.18135837  & $-$0.18057175&0.1826114098&\\

2  &3  & $-$0.18211323 &  $-$0.18086323  & $-$0.17991337 &  $-$0.17833885  & $-$0.17676468&0.180993962&\\

2  &4  & $-$0.18061374 &  $-$0.17853041  & $-$0.17694447 &  $-$0.17431782  & $-$0.17169211&0.178847328&\\\hline
\end{tabular}
\begin{tabbing}
	\footnotemark[1]{These energies compare with those from \cite{shd.ijmpa2008}}. \ \ \ \ \ \ \  \=
	\footnotemark[2]{These are calculated here using the GPS method, for the first time.}  \\ 
\end{tabbing} 
\end{table} 

\end{document}